\documentclass[aps,prb,twocolumn,showpacs,preprintnumbers,amsmath,amssymb,superscriptaddress,longbibliography,nofootinbib]{revtex4-2}
\pdfoutput=1
\usepackage{hyperref}
\hypersetup{colorlinks=true, citecolor=black,  linkcolor=black, urlcolor=black}
\usepackage{color}
\usepackage{graphicx}
\usepackage{dcolumn}
\usepackage{braket}
\usepackage{amsthm}
\usepackage[caption=false]{subfig}

\newcommand{\ex}[1]{\mathrm{e}^{#1}}
\newcommand{\fr}{\frac}
\newcommand{\pa}[1]{\left(#1 \right)}

\newcommand{\bb}[1]{\mathbb{#1}}
\newcommand{\abs}[1]{\left|#1\right|}

\def\tr{{\mathrm{Tr}}}
\def\del{{\partial}}

\makeatletter
\let\cat@comma@active\@empty
\makeatother

\begin{document}
\title{Universality of Rényi Entropy in Conformal Field Theory}
\preprint{CALT 2025-008}
\preprint{IPMU 25-0015}
\preprint{KYUSHU-HET-314}
\preprint{RIKEN-iTHEMS-Report-25}

\author{Yuya Kusuki}\email[]{\textcolor{black}{kusuki.yuya@phys.kyushu-u.ac.jp}}
\affiliation{\it Institute for Advanced Study, 
Kyushu University, Fukuoka 819-0395, Japan.}
\affiliation{\it Department of Physics, 
Kyushu University, Fukuoka 819-0395, Japan.}
\affiliation{\it RIKEN Interdisciplinary Theoretical and Mathematical Sciences (iTHEMS),
Wako, Saitama 351-0198, Japan.}

\author{Hirosi Ooguri}\email[]{\textcolor{black}{ooguri@caltech.edu}}
\affiliation{\it Walter Burke Institute for Theoretical Physics, California Institute of Technology, Pasadena, CA 91125, USA}
\affiliation{\it Kavli Institute for the Physics and Mathematics of the Universe (WPI), University of Tokyo, Kashiwa 277-8583, Japan}

\author{Sridip Pal}\email[]{\textcolor{black}{sridip@caltech.edu}}
\affiliation{\it Walter Burke Institute for Theoretical Physics, California Institute of Technology, Pasadena, CA 91125, USA}

\begin{abstract}
We use the thermal effective theory to prove 
that, for the vacuum state in any conformal field theory in $d$ dimensions, the $n$-th Rényi entropy $S_A^{(n)}$ behaves as
$S_A^{(n)} = \frac{f}{(2\pi n)^{d-1}} \frac{ {\rm Area}(\partial A)}{(d-2)\epsilon^{d-2}}
\left(1+O(n)\right)$ in the $n \rightarrow 0$ limit  
when the boundary of the entanglement domain $A$ is spherical with the UV cutoff $\epsilon$.
The theory dependence is encapsulated in the cosmological constant $f$ in the thermal effective action. Using this result, we estimate the density of states for large eigenvalues of the modular Hamiltonian for the domain $A$.  In two dimensions, we can use the hot spot idea to derive more powerful formulas valid for arbitrary positive $n$. We discuss the difference between two and higher dimensions and clarify the applicability of the hot spot idea.
We also use the thermal effective theory to derive an analog of the Cardy formula for boundary operators in higher dimensions.

\end{abstract}
\maketitle
\section{Introduction and Summary}

In conformal field theory (CFT) there are universal formulas that describe how physical quantities depend on the theory. Arguably, the most famous example is the Cardy formula for {\it 2d} CFT \cite{Cardy1986}, which estimates the density of states $\rho(\Delta)$ as a function of the conformal dimension $\Delta$ as,
\begin{equation}\begin{aligned}
&\int_{\Delta - \delta}^{\Delta+\delta} d\Delta' \rho(\Delta') \\
& = \left(\frac{c}{48\Delta^3}\right)^{1/4} \exp\left(2\pi \sqrt{\fr{c}{3} \Delta}+O(1) \right),  
\label{eq:Cardy}
\end{aligned}
\end{equation}
where $c$ is the central charge of the theory. The formula is valid for $\Delta \gg c$ if we choose $\delta=O(1)$ \cite{Mukhametzhanov:2019pzy,Mukhametzhanov:2020swe}.

The method of thermal effective theory introduced in \cite{Bhattacharyya:2007vs,Banerjee:2012iz,Jensen:2012jh} and further developed in \cite{Benjamin2023,Kang2023} turned out to have powerful applications to many problems in CFT including the generalization of the Cardy formula to higher dimensions \cite{Bhattacharyya:2007vs, Shaghoulian:2015lcn}, the supersymmetric indices \cite{DiPietro:2014bca}, the asymptotic properties of $($heavy$)^3$ OPE coefficients \cite{Benjamin2023}, the internal and spatial symmetry resolutions of partition functions \cite{Kang2023, Benjamin2024}, their subleading terms \cite{Allameh:2024qqp}, and effects of boundaries and defects \cite{Diatlyk2024, Kravchuk:2024qoh, Cuomo:2024psk, Banihashemi:2025qqi}. Compactifying
$d$-dimensional CFT on the thermal circle $\bb{S}^1_\beta$ at temperature $\beta^{-1}$,  the high temperature limit 
$\beta^{d-1} \ll {\rm vol}(\Sigma_{d-1})$
is described by a quantum field theory 
on $\Sigma_{d-1}$ with a mass gap $m_{\text{gap}} = 1/\beta$. It is convenient to use the Kaluza-Klein (KK) ansatz for the $d$-dimensional metric $G_{\mu\nu}$ on $\bb{S}^1_\beta \times \Sigma_{d-1}$ as,
\begin{equation}\label{eq:KKmetric}
\begin{aligned}
G_{\mu \nu} dx^\mu dx^\nu
&=g_{ij}(\vec{x}) dx^i dx^j + \ex{2\phi(\vec{x})}\pa{ d\tau + A_i(\vec{x})dx^i  }^2 \\
&=\ex{2\phi(\vec{x})} \left[   \hat{g}_{ij}(\vec{x}) dx^i dx^j + \pa{ d\tau + A_i(\vec{x})dx^i  }^2 \right],
\end{aligned}
\end{equation}
where $\tau$ is the Euclidean time with the period $1$ and $\hat{g}_{ij} = e^{-2\phi}g_{ij}$  is the Weyl-invariant metric on $\Sigma_{d-1}$. Due to the conformal invariance in $d$ dimensions, the thermal partition function and related physical observables depend on the KK scalar $\phi$ only through the Weyl anomaly \cite{Benjamin2023} and does not contribute to the leading terms in the $\beta \rightarrow 0$ limit, which we study in this paper. Thus, we can safely ignore the $\phi$ dependence. The rest of 
the low energy effective action can be expanded in
the power of derivatives as
\begin{equation}\label{eq:th}
S[\hat{g},A] = \int d^{d-1}x \sqrt{\hat{g}} \pa{ -f + c_1 \hat{R} + c_2 F^2 + \cdots  },
\end{equation}
where 
$\hat{R}$ is the scalar curvature for the metric $\hat{g}_{ij}$, and $F_{ij}$ is the field strength of the KK gauge field, $F_{ij} = \partial_iA_j- \partial_jA_i $. The first term measures the volume of $\Sigma_{g-1}$. When $A_i=0$, it gives
\begin{equation}
    \int d^{d-1} \sqrt{\hat g} = \beta^{1-d}
    \  {\rm vol}(\Sigma_{d-1}).
\end{equation}
The cosmological constant $f$ multiplying to it has several important physical interpretations as explained in \cite{Benjamin2023}. When $d=2$, 
 the Cardy formula (\ref{eq:Cardy}) shows that $f$ is given by the central charge $c$ as,
\begin{equation}
f=\fr{2\pi c}{12}.
\end{equation}

The thermal effective theory is also useful when the temperature $\beta^{-1}$ is 
spatially dependent. Under certain conditions,
which we will clarify in this paper, 
leading contributions to the thermal effective action come from the regions where the local temperature is high, and we only need to evaluate the integral of the Lagrangian density in these regions. 
In \cite{Benjamin2023}, this ``hot spot idea" was developed and successfully applied to derive the universal formula for OPE coefficients. 

In this paper, we use the thermal effective theory to study properties of the R\'enyi entropy defined by
\begin{equation}
S^{(n)}_A = \fr{1}{1-n} \log \tr \rho_A^n,
\end{equation}
where $\rho_A$ is the reduced density matrix for a subsystem $A$. 
In \cite{Agon2023},  
it was conjectured that the Rényi entropy has the universal behavior in the $n \rightarrow 0$ limit as
\begin{equation}
 S^{(n)}_A \sim \frac{f}{n^{d-1}} \ g(A), 
\label{universality0}
\end{equation}
where the dependence of the CFT is encapsulated in the cosmological constant $f$, and
$g(A)$ depends only on the geometry of the subsystem $A$.   In this paper, we recast the argument in Section 3.1 of \cite{Agon2023} in the language of the thermal effective theory, which allows us to justify their use of the position-dependent temperature, to take into accout boundary and curvature effects, and to estimate subleading terms in the small $n$ expansion.

In Section \ref{sec:uni} of this paper, we  verify the universal behavior \eqref{universality0} when the boundary of $A$ is spherical and determine $g(A)$ when $\epsilon^{d-2} \ll {\rm Area}(\partial A)$ as,
\begin{equation}
    g(A) =\frac{{\rm Area}(\partial A)}{(2\pi )^{d-1}(d-2)\epsilon^{d-2}} .
    \label{sphereexpression}
\end{equation}
In Section \ref{sec:surface}, we generalize the thermal effective theory to include effects of boundaries  
 and estimate subleading terms to eq. \eqref{universality0} in the small $n$ expansion as
\begin{equation}
 S^{(n)}_A =\frac{f}{n^{d-1}} \ g(A)
+O\left(\frac{1}{n^{d-2}}\right), ~~~ n \ll 1.
\label{universality}
\end{equation}

As an application of the universal formula (\ref{universality}),
 we study the entanglement spectrum 
$d_\lambda$ defined by
\begin{equation}
    S_A^{(n)}  =
\sum_\lambda d_\lambda \ex{-2\pi n \lambda} ,
\end{equation} 
in Section \ref{sec:ES}.
For $\lambda \gg 1$,
we estimate 
$d_\lambda$ for large $\lambda$ as,
\begin{equation}
\begin{aligned}
&d_\lambda \ {\sim} \ 
\frac{1}{\sqrt{2\pi d}}
\left(\frac{d-1}{d-2}\frac{f}{\lambda^{d+1}} \frac{\text{Area}(\del A)}{\epsilon^{d-2}} \right)^{\frac{1}{2d}}\\
&\times
\exp\left[ \frac{d}{(d-2)^{\frac{1}{d}}(d-1)^{1-\frac{1}{d}}} \left( \frac{f \lambda^{d-1} \text{Area}(\del A)}{\epsilon^{d-2}}
\right)^{\frac{1}{d}}
\right] \, .\\
\end{aligned}
\end{equation}

In two dimensions, we can use the hot spot idea to derive a result stronger than the combination of eqs. \eqref{sphereexpression} and \eqref{universality}, which is applicable to arbitrary  $n$ as,
\begin{equation}
S^{(n)}_A = \fr{c}{6}\pa{\fr{1}{n}+1}\log \fr{\abs{A}}{2\epsilon} + s_a + s_b,
\label{eq:Sn}
\end{equation}
where we followed the prescription of
\cite{Ohmori2015,Cardy:2016fqc} to impose conformal boundary conditions at the end points $a$ and $b$ of the subsystem $A$ (see Fig. \ref{fig:EC})
and 
$s_a = \log \braket{0|a}$ is the boundary entropy of $a$, introduced in \cite{Affleck1991}. This formula was originally found by using a conformal map to relate $S^{(n)}_A$ to the annulus partition function \cite{Cardy:2016fqc}. The leading term in \eqref{eq:Sn} can also be captured using the twist field approach \cite{Calabrese2004}.

One might expect a similar formula for arbitrary  $n$ in higher dimensions. 
However, it turns out that the hot spot idea is not directly applicable in such a case because of the behavior of the effective temperature $\beta^{-1}$ near the boundary of the entanglement domain, as we will explain in Section \ref{sec:nonuni}. This does not necessarily mean that eq. \eqref{eq:Sn} does not generalize to higher dimensions, but it suggests that the hot spot idea has a certain limitation.
We discuss the necessary conditions for the applicability of the hot spot idea and show that it can be used consistently for  
the computation in \cite{Benjamin2023} and that the universal formula for OPE coefficients derived there using the hot spot idea remains valid.

As another application of the thermal effective theory, in Section \ref{sec:BdySpec} we generalize the Cardy formula to higher dimensional boundary conformal field theories (BCFTs).
We end this paper with a discussion on future directions of research in Section \ref{sec:discussion}.

\bigskip

\noindent
{\bf Note Added:} Towards the completion of this work, we received \cite{Agon2025}, where the conjecture (\ref{universality0}) was tested for holographic CFTs using the AdS/CFT correspondence. The thermal effective theory approach in this paper is more general and applicable to any CFT, whether or not it is unitary or has a holographic description, as far as the R\'enyi entropy is well-defined.

\section{Universality of Rényi Entropy}\label{sec:uni}
Consider the $n$-th Rényi entropy for a spherical entanglement domain $A$ of radius $R$ in a $d$-dimensional CFT on Minkowski space $\mathbb{R}^{1,d-1}$ with the following metric,
\begin{equation}
ds^2 = -dt^2 + dr^2 +r^2 d\Omega^2_{d-2},
\end{equation}
where $d\Omega_{d-2}$ is the metric of the unit transverse sphere $\bb{S}^{d-2}$.
To calculate the Rényi entropy using the thermal effective theory,  
it is useful to consider the following coordinate transformation \cite{Casini:2011kv},
\begin{equation}
\begin{aligned}
t &= R \fr{\sinh(\tau/R) }{  \cosh u + \cosh(\tau/R)     },\\
r &= R \fr{   \sinh u   }{  \cosh u + \cosh(\tau/R)   }.
\end{aligned}
\end{equation}
After the Euclidean rotation, $\tau \rightarrow -i \tau$, the metric takes the following form,
\begin{equation}
ds^2 = \Xi^2 \pa{ d\tau^2  + R^2(du^2 + \sinh^2 u \ d\Omega_{d-2}^2)   },
\end{equation}
where $\Xi = \pa{\cosh u + \cos(\tau/R)}^{-1} $
and the periodicity in $\tau$ is $2 \pi R n$.
By removing the factor $\Xi$ by the Weyl invariance and rescaling $\tau$ so that its period is $1$,  
the CFT is defined on $\bb{R} \times \bb{H}^{d-1}$ with the metric in the KK form  (\ref{eq:KKmetric}),
\begin{equation}\label{eq:metric}
ds^2 =  d\tau^2  + \fr{1}{(2\pi n)^2}\pa{du^2 + \sinh^2 u d\Omega_{d-2}^2} .
\end{equation}

To define the reduced density matrix,  
it is necessary to decompose the Hilbert space of the CFT into the one over $A$ and the one over its complement $\bar A$.
Here, we perform the decomposition by introducing a physical boundary between the subsystem $A$ and $\bar{A}$, placing the cutoff surface at $r=R-\epsilon$.
In the hyperbolic coordinates, 
this cutoff surface is located at $u=\log (2R/\epsilon)$ in $\bb{H}^{d-1}$, to the leading order in $(R/\epsilon)$.
In other words, the 
cutoff corresponds to a regularization of the infinite volume of $\mathbb{H}^{d-1}$.  

We now consider the thermal effective theory on  $\bb{S}^1_\beta \times \mathbb{H}^{d-1}$.  
Since the mass gap $\beta^{d-1}$ 
is much smaller than the regularlized volume
of $\mathbb{H}^{d-1}$,  
the thermal effective theory can be described by a local action.
In particular, we can calculate the Rényi entropy 
by the thermal effective action (\ref{eq:th}).  
Here, the inverse temperature
$\beta$
is proportional to the replica number $n$.  
Expanding in the power of $n$,  
the leading term is 
\begin{equation}\label{eq:bulk-leading}
 -f \int d^{d-1}x \sqrt{\hat{g}} =
  \frac{f}{(2\pi n)^{d-1}} \frac{ {\rm Area}(\partial A)}{(d-2)\epsilon^{d-2}},
\end{equation}
in the limit of $n\rightarrow 0$ and then $\epsilon^{d-2} \ll {\rm Area}(\partial A) $,   and the subleading bulk term 
\begin{equation}
\label{eq:bulk-sub-leading}
 c_1 \int d^{d-1}x \sqrt{\hat{g}} \hat{R} = -\frac{(d-1) c_1}{(2 \pi n)^{d-3}} \frac{{\rm Area}(\partial A)}{\epsilon^{d-2}}    ,
\end{equation}
is suppressed by $n^2$ compared to the leading term.
In these and later equations, we only exhibit the leading terms in $(R/\epsilon)$ since $\epsilon$ is the UV cutoff.
Note that the powers of $(R/\epsilon)$ are the same for both, which will become important when we discuss the applicability of the hot spot idea in Section \ref{sec:nonuni}. 

There are also contributions from the boundary at $u=\ln (2R/\epsilon)+O(\epsilon)$.
As we will see in Section \ref{sec:surface}, the leading boundary contribution is of the order $1/n^{d-2}$, suppressed by one power of $n$ compared to the leading bulk term \eqref{eq:bulk-leading}.
Thus, the $n \to 0$ limit of the Rényi entropy is given by the following universal formula including the estimate of the leading correction,
\begin{equation}\label{eq:mainResult}
S_A^{(n)} = \frac{f}{(2\pi n)^{d-1}} \frac{ {\rm Area}(\partial A)}{(d-2)\epsilon^{d-2}}
 +O\left(
\frac{1}{n^{d-2}}\right).
\end{equation}

So far, we have considered the case when the boundary of $A$ is spherical. It is natural to conjecture that eq. \eqref{eq:mainResult}
generalizes to arbitrary shapes of $A$ as far as it is connected.
In Section \ref{sec:discussion}, we will discuss a possible way to prove this conjecture.

In two dimensions, the polynomial divergence  
in (\ref{eq:bulk-leading}) is replaced by a logarithmic divergence,  
\begin{equation}\label{eq:2dEE}
 -f \int d^{d-1}x \sqrt{\hat{g}} = - \fr{f}{ \pi n }   \log \fr{|A|}{2\epsilon} = -\fr{c}{ 6n }   \log \fr{|A|}{2\epsilon}.
\end{equation}
Including the estimate of the boundary terms, we obtain
\begin{equation}
S^{(n)}_A =\fr{c}{ 6n }   \log \fr{\abs{A}}{2\epsilon} + O(1), ~~~ n \ll 1.
\end{equation}
This reproduces with the $n \to 0$ limit of the universal formula (\ref{eq:Sn}) for the Rényi entropy in two dimensions.
Note that the denominator appears as $2\epsilon$ instead of $\epsilon$,  
because to match the convention used in higher dimensions,
we follow the convention shown in Fig.\ref{fig:EC},  
where the boundary size is taken to be $2\epsilon$ rather than $\epsilon$, which is the standard convention in {\it 2d} CFT.  

As a test of the universal formula \eqref{eq:mainResult}, let us consider a holographic CFT,
where the cosmological constant $f$ in the thermal effective action is given \cite{Benjamin2023} as 
\begin{equation}\label{eq:f_holo}
f = \fr{(4\pi l_{\mathrm{AdS}})^{d-1}}{4G_N
d^d },
\end{equation}
where $l_{\mathrm{AdS}}$ is the AdS radius,  
and $G_N$ is the $d+1$-dimensional Newton constant.
Thus, the $n\to 0$ limit of the Rényi entropy is given by
\begin{equation}
S^{(n)}_A =
 \pa{\fr{2 l_{\mathrm{AdS}}}{n}}^{d-1}    \fr{\text{Area}(\del A)}{ 4G_N d^d(d-2)\epsilon^{d-2}}  
 + O\left(\frac{1}{n^{d-2}}\right) , 
\end{equation}
This result agrees with the holographic computation of Rényi entropy based on \cite{Hung2011, Agon2025}.
This can also be regarded as a consistency check of (\ref{eq:f_holo}) (We can perform a scheme independent check by comparing coefficients of the logarithmic divergences in even dimensions.)  \\

\section{Contribution from Boundary of Entanglement Domain}\label{sec:surface}

\begin{figure}[t]
 \begin{center}
  \includegraphics[width=4.0cm,clip]{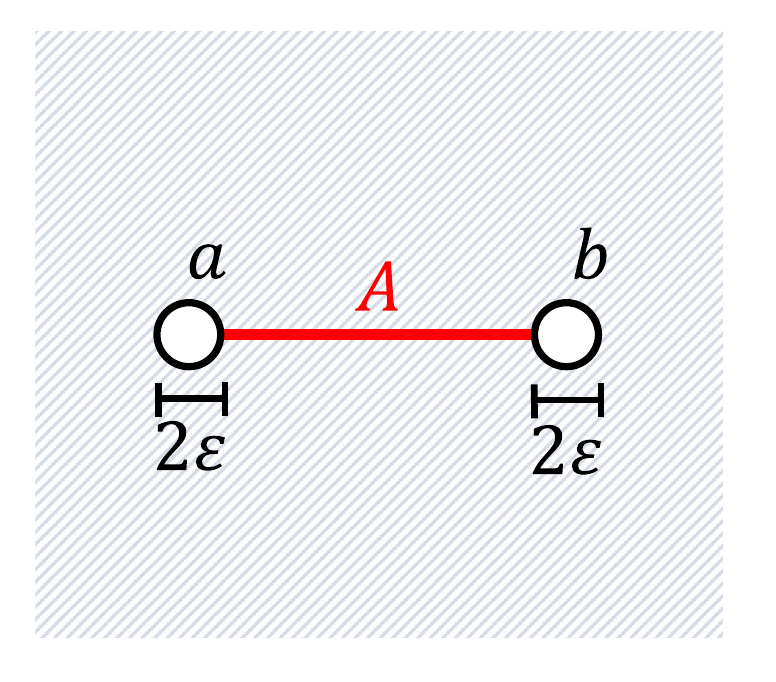}
 \end{center}
 \caption{Regularization prescription in {\it 2d} CFT:  By inserting a physical boundary of size $2\epsilon$ at the entangling surface,  
we decompose the CFT Hilbert space.
}
 \label{fig:EC}
\end{figure}

To regularize the calculation of the R\'enyi entropy, we insert a physical boundary between the entanglement domain
$A$ and its complement $\bar A$.  First, we need to specify the boundary condition.
In {\it 2d} CFT, it has been studied in \cite{Ohmori2015,Cardy:2016fqc}. Assuming that   
the boundaries $a$ and $b$ at the two end-points of $A$ (as shown Fig. \ref{fig:EC}) are conformal, preserving
one half of the bulk conformal symmetry,
the boundary contributions to 
the Rényi entropy (\ref{eq:Sn}) are given by
\begin{equation}
S^{(n)}_A = \fr{c}{6}\pa{1+\fr{1}{n}}\log \fr{\abs{A}}{2\epsilon} + s_a + s_b,
\end{equation}
where $s_a = \log \braket{0|a}$ is the boundary entropy of $a$, introduced in \cite{Affleck1991}.

To generalize this to higher dimensions, we need to
take into account boundary terms in 
the thermal effective action,
\begin{equation}\label{eq:Sfull}
\begin{aligned}
S[\hat{g},A]  &= \int d^{d-1}x \sqrt{\hat{g}} \pa{ -f + c_1 \hat{R} + c_2 F^2 + \cdots  } \\
&+\int d^{d-2}x \sqrt{\hat{\sigma}} \pa{ -g + b_1 \hat{K} + b_2 \hat{R} + \cdots  },
\end{aligned}
\end{equation}
where $\sigma$ is the induced metric on the boundary and $K$ is the trace of the extrinsic curvature $K_{ab} =\nabla_a n_b$ where $n$ is a unit vector normal to the boundary.
The leading contribution from the boundary effective action is
\begin{equation}\label{eq:boundary-leading}
 -g \int d^{d-2}x \sqrt{\hat{\sigma}} = -\frac{g}{n^{d-2}} \frac{{\rm Area}(\partial A)}{ (\pi\epsilon)^{d-2}  }   .
\end{equation}
This is suppressed by one power of $n$ relative to the bulk leading term \eqref{eq:bulk-leading} as indicated in \eqref{eq:mainResult}.  

Let us examine the boundary leading contribution (\ref{eq:boundary-leading}) explicitly in a {\it 2d} CFT.
Unlike the computation of the bulk leading contribution (\ref{eq:2dEE}),  
the boundary leading contribution is given by a zero-dimensional integral, so no logarithmic divergence appears.  
Thus, the contribution of the boundary to the Rényi entropy is simply given by a constant $2g$.  
Here, the factor of 2 accounts for the fact that the entangling surface is disconnected in {\it 2d} CFT.  

In two dimensions, there are only two local terms one can write using Weyl-invariant metric, as follows: 
\begin{equation}
\begin{aligned}
S[\hat{g},A]  &= -f \int d^{d-1}x \sqrt{\hat{g}}  -  2g\int d^{d-2}x \sqrt{\hat{\sigma} } .
\end{aligned}
\end{equation}
Thus, we obtain in the $\epsilon \to 0$ limit for arbitrary  $n$, 
\begin{equation}
S^{(n)}_A = \fr{c}{6}\pa{1+\fr{1}{n}}\log \fr{\abs{A}}{2\epsilon} + 2g.
\end{equation}
In this expression, $g$ plays the role of the boundary entropy.  
Investigating $g$ in various examples of higher-dimensional CFTs  
is an interesting future direction.  
In particular, since the holographic dual  
of the lattice regularization introduced here  
is provided in \cite{Kusuki2023},  
it would be intriguing to use this framework  
to compute $g$ in holographic CFTs.  

\section{Entanglement Spectrum}\label{sec:ES}

The density matrix $\rho$ can be written in terms of modular Hamiltonian $K$:
\begin{equation}
   \rho= \frac{\ex{-2\pi K}}{\tr\ \ex{-2\pi K}}
\end{equation}
The eigenvalues of $\rho$ lie in $[0,1]$. Consequently, the eigenvalues of $K$ lie in $[0,\infty)$. We would like to estimate the density/degeneracy of low-lying eigenvalues of $\rho$. They correspond to the large eigenvalues of $K$. See \cite{Calabrese:2008iby,Alba:2017bgn,Kusuki2023a} for the computation of the density of large eigenvalues of $K$ in two dimensions and \cite{Baiguera:2024ffx} for the holographic computation of the same in higher dimensions.

We recall the derivation of the Cardy formula for the high-energy density of states of a CFT on spatial slice $\bb{S}^{d-1}$,
from the leading term in the $\beta\to 0$ expansion
of the partition function of the CFT on the manifold $\bb{S}^1_\beta\times \bb{S}^{d-1}$; see, for example, \cite{Cardy1986,Benjamin2023, Kang2023, Benjamin2024, Diatlyk2024, Kravchuk:2024qoh, Cuomo:2024psk}. In other words, the expansion of $\tr\ \ex{-\beta H}$ in $\beta$ provides the density of large eigenvalues of $H$.
Now that we have identified the leading term
in the $n$ expansion of the $n$-th Rényi entropy i.e. we know $\tr\ \ex{-2\pi n K}$ in the $n\to 0$ limit, 
we can determine the entanglement spectrum asymptotically, i.e. the large eigenvalues of $K$,  
in the same spirit as the Cardy formula. In particular, the \eqref{eq:mainResult} can be written as: 
\begin{equation}\label{eq:approx}
\begin{aligned}  
S_A^{(n)} & =
\sum_\lambda d_\lambda \ex{-2\pi n \lambda} \\
& = \exp\left[ \fr{f}{(2\pi n)^{d-1}} \fr{\text{Area}(\del A)}{ (d-2) \epsilon^{d-2} }   + O\left(\frac{1}{n^{d-2}}\right)\right] \,.
\end{aligned}
\end{equation}
Here the sum over $\lambda$ 
on the right-hand side of the first line runs over the eigenvalues $\lambda$ of $K$ and $d_\lambda$ is the degeneracy of the eigenvalue $\lambda$. 

The inverse Laplace transform gives 
\begin{equation}
    \begin{aligned}       d_\lambda &= \frac{1}{2\pi i}\int_{\gamma-i\infty}^{\gamma+i\infty} dy\ \left(\sum_{\lambda'} d_{\lambda'} \ex{-y \lambda'} \right) \ex{y\lambda} \\
        &= \frac{1}{2\pi i}\int_{\gamma-i\infty}^{\gamma+i\infty} dy\ 
   S_A^{(n=\frac{y}{2\pi})}  \ex{y\lambda}   .
        \,
    \end{aligned}
\end{equation}
Because of the factor $\ex{y \lambda}$, the large $\lambda$ asymptotics of $d_\lambda$ is expected to come from the region near $y=0$ in the saddle point approximation, where we can use \eqref{eq:approx} for $S_A^{(n=\frac{y}{2\pi})}$.
 The saddle is at 
$$\lambda_*=f\fr{(d-1) \text{Area}(\del A)}{(d-2)\epsilon^{d-2}}y^{-d}\,$$ 
and gives us with the estimate for $d_\lambda$ for $\lambda \gg 1$ as,
\begin{equation}
\begin{aligned}\label{eq:naive}
&d_\lambda \ {\sim} \ 
\frac{1}{\sqrt{2\pi d}}
\left(\frac{d-1}{d-2}\frac{f}{\lambda^{d+1}} \frac{\text{Area}(\del A)}{\epsilon^{d-2}} \right)^{\frac{1}{2d}}\\
&\times
\exp\left[ \frac{d}{(d-2)^{\frac{1}{d}}(d-1)^{1-\frac{1}{d}}} \left( \frac{f \lambda^{d-1} \text{Area}(\del A)}{\epsilon^{d-2}}
\right)^{\frac{1}{d}}
\right] \, .\\
\end{aligned}
\end{equation}
We can express this limit more precisely if we assume analyticity in $n$.
In particular, we can use Ingham's Theorem \cite{ingham1941tauberian} (see Theorem IV.21.1 in \cite{korevaar2004tauberian} for a textbook exposition) to obtain the following:
\begin{equation}
\begin{aligned}
 &\sum_{\lambda=0}^{\Lambda} d_\lambda
 = \frac{1}{\sqrt{2\pi}}
 \left(\frac{d-2}{(d-1)d^{d-1}}
 \fr{\epsilon^{d-2}}{f\Lambda^{d-1} \text{Area}(\del A) }\right)^{\fr{1}{2d}} 
\\
&\times
\exp\left[ \frac{d}{(d-2)^{\frac{1}{d}}(d-1)^{1-\frac{1}{d}}} \left(  \frac{f\Lambda^{d-1}\text{Area}(\del A)}{\epsilon^{d-2}}
\right)^{\frac{1}{d}} \right. \\
& ~~~~~~~~~~~~~ \left.
+ ~ O\left(\Lambda^{\fr{d-2}{d}}\right) \right]
\times \left(1+o(1)\right).
\end{aligned}
\end{equation}
This is consistent with the naive formula \eqref{eq:naive} for the asymptotic spectrum when integrated against a window function supported on $[0,\Lambda]$ in the $\Lambda\to\infty$ limit. A similar result has been obtained in the context of the Cardy formula in Appendix C of \cite{Das:2017vej} and in \cite{Mukhametzhanov:2019pzy}.

\section{Applicability of Hot Spot Idea}\label{sec:nonuni}

In two dimensions,
the universal formula (\ref{eq:Sn}) for the Rényi entropy 
originates from the $\epsilon \to 0$ limit, rather than the $n \to 0$ limit. That is why the formula is applicable to arbitrary $n$ not just in the $n \rightarrow 0$ limit.
Can we find a similar formula in higher dimensions?  

Let us consider the following coordinates,
\begin{equation}
ds^2 =   d\tau^2  + \fr{1}{(2\pi n \sin\theta)^2}\pa{d\theta^2 + \cos^2 \theta d\Omega_{d-2}^2},
\end{equation}
where $0\leq \theta \leq \pi/2$.
They are related to the coordinates (\ref{eq:metric}) by the coordinate transformation $\sinh u = \cot \theta$ with a conformal transformation removing a conformal factor.
This metric shows the spatial
dependence $\beta(\theta) \sim \sin \theta$ of the temperature $1/\beta$, which diverges linearly 
at $\theta = 0$.
It was suggested in \cite{Benjamin2023} that,  
when the temperature has spatial dependence and diverges at some points, 
the thermal effective action is dominated by contributions from the regions near the divergence, which they called the hot spot. 

In the current setup, the hot spot is expected to be in the vicinity of $\theta = 0$ where
$\beta = 0$.
In order for the hot spot idea to work,  
the higher-order terms in the effective action must be suppressed in the region.
With the spatial dependence of the  temperature given by,
\begin{equation}
\beta^2 = \ex{-2\phi(\vec{x})},
\end{equation}
where $\phi(\vec{x})$ is given by (\ref{eq:KKmetric}),
the Ricci scalar $\hat{R}$ is evaluated as
\begin{equation}
\begin{aligned}
\hat{R} &= \beta^2 R + 2(d-2)  \beta^2\nabla^2\log \beta \\
&+ (d-3)(d-2) \beta^2\del_a \log \beta \del^a \log \beta .
\end{aligned}
\end{equation}
Therefore, higher order terms are suppressed only if the logarithmic derivatives of $\beta$ do not diverge too quickly.
Indeed, this is the case for the scalar curvature in two dimensions because of the $(d-2)$ factor in the second and third terms on the right-hand side in the above equation.
However, this is not the case in higher dimensions.
For example, in three dimensions, $\nabla^2\log \beta $ diverges as $\beta^{-2}$
at the hot spot, and the scalar curvature remains finite $\hat{R} = -6$. Thus, 
the higher-order terms contribute in the same order  
as the cosmological constant term, and the high temperature expansion of the thermal effective action fails near the hot spot.

This is related to the fact that both the cosmological term \eqref{eq:bulk-leading} and the Einstein-Hilbert term \eqref{eq:bulk-sub-leading} in the thermal effective action have the same power in $(R/\epsilon)$, as we observed earlier.
On the other hand, they have different powers in $n$, which is why the small $n$ expansion of the thermal effective action is valid.

The fact that the leading contribution  
to the R\'enyi entropy for arbitrary   $n$ can be universally expressed  
in terms of the central charge,  
seems to be a unique property in two dimensions.
Our result suggests that, in higher dimensions, the R\'enyi entropy for general $n$ depends  on the details of the theory (See \cite{Belin:2013dva} for related comments in the context of holography). 

In contrast, in the set-up of the paper \cite{Benjamin2023}, $\beta$
vanishes quadratically in the coordinate corresponding to $\theta$, and $\beta^2 \nabla^2 \log \beta$ and $\beta^2 \partial_a \log \beta \partial^a \log \beta$ vanishes at the hot spot. Therefore, the hot spot idea can be consistently applied there, and the universal OPE formula discovered in the paper remains unaffected by the subtlety we point out here.

\section{Asymptotic Density of Boundary States}\label{sec:BdySpec}

In two dimensions, the comparison of the open and closed string channels for the annulus zero-point function gives the following asymptotic formula for the density of states $\rho(\Delta)$ in the open string Hilbert space with the
boundary condition $a$,
\begin{equation}\label{eq:open}
\begin{aligned}
&\int_{\Delta-\delta}^{\Delta+\delta}d\Delta'  \rho(\Delta') \\
&= \left(\frac{c}{96\Delta^3}\right)^{1/4}\exp\left(2\pi \sqrt{\fr{c}{6} \Delta} + 2 s_a+O_\delta(1)\right)\,,
\end{aligned}
\end{equation}
The above formula is valid in the regime $\Delta\gg c$ and $\delta=O(1)$. 
The subscript in $O_\delta(1)$ indicates that the error is in the range bounded by $\left[\log(2\delta-1),\log(2\delta+1)\right]$, which depends on $\delta>1/2$ \cite{Kusuki2023a}.

To generalize this to higher dimensions, consider the 
CFT on $\mathbb{S}^1_\beta \times \mathbb{S}_+^{d-1}$, where $\mathbb{S}_+^{d-1}$ is a hemisphere
with a $\mathbb{S}^{d-2}$ boundary.
The first few terms in the $\beta$ expansion of the effective action are given by
\begin{equation}
\begin{aligned}
S[\hat{g}, A] &= -f \fr{\mathrm{vol}(\bb{S}^{d-1})}{2\beta^{d-1}} - g \fr{\mathrm{vol}(\bb{S}^{d-2})}{\beta^{d-2}} \\
&+ c_1 \fr{(d-2)(d-1)\mathrm{vol}(\bb{S}^{d-1})}{2\beta^{d-3}} + \cdots.
\end{aligned}
\end{equation}
By performing the inverse Laplace transform with the saddle point approximation,  
we obtain the higher-dimensional generalization of (\ref{eq:open}),
\begin{equation}
\begin{aligned}
\log \rho(\Delta) = f^{\frac{1}{d}}\cdot \fr{d}{d-1}&\pa{\fr{(d-1){\rm vol}(\bb{S}^{d-1})}{2}}^{\fr{1}{d}}
\Delta^{\fr{d-1}{d}} \\
+f^{\frac{2-d}{d}} g \cdot {\rm vol}(\bb{S}^{d-2}) &\pa{\fr{(d-1)\mathrm{vol}(\bb{S}^{d-1})}{2}}^{\fr{2-d}{d}}
\Delta^{\fr{d-2}{d}} \\
&+ O(\Delta^{\fr{d-3}{d}}).
\end{aligned}
\end{equation}

\section{Discussion}\label{sec:discussion}
In this paper, we use thermal effective theory to find universal properties of the R\'enyi entropy in the limit of $n\rightarrow 0$. Our results suggest several future research directions:
\begin{itemize}
\item In this paper, we discussed how the thermal effective action  is modified in the presence of spatial boundaries and used 
the result to estimate the R\'enyi entropy, the entanglement spectrum, and the asymptotic density of boundary states.   
We expect many other applications of the thermal effective theory to BCFT. An interesting direction may be to use it to generalize the known universality in two dimensions   
\cite{Kusuki2022, Numasawa2022} to higher dimensions.

\item In this paper, we restricted our attention to the case when the boundary of the entanglement domain $A$ is spherical. It would be interesting to see if the universal formulas such as eq. \eqref{universality} generalize to arbitrary shapes of $A$ as far as it is connected. A possible approach may be to deform the Weyl invariant metric $\hat{g}_{ij}$ and use the thermal effective action to understand the responses of the R\'enyi entropy and
other physical quantities. 

\item If CFT has a global symmetry $G$, we can turn on the gauge field and use the method developed in 
 \cite{Casini:2019kex, Magan:2021myk, Kang2023} to obtain the symmetry resolution of the Rényi entropy \cite{Belin:2013uta,Goldstein:2017bua,xavier2018equipartition,Calabrese:2021wvi}, as 
\begin{equation}\label{eq:SREE}
\lim_{n \rightarrow 0}
\left( S^{(n)}_A(r)- S^{(n)}_A \right) = \log \fr{d_r^2}{\abs{G}},
\end{equation}
where $S^{(n)}_A(r)$ is the $n$-the Rényi entropy within the irreducible representation $r$, $d_r$ is the quantum dimension of $r$, and $\abs{G}$ is the order of the group $G$.
Our result \eqref{eq:SREE}  is the first step towards the higher dimensional generalization of the results for {\it 2d} CFTs \cite{Kusuki2023a}, where it was shown that 
the equation holds for arbitrary $n$.
\end{itemize}

\bigskip

\section*{Acknowledgments}
The authors thank David Simmons-Duffin for helpful discussions and comments on the draft. This research was  supported in part
by the U.S. Department of Energy, Office of Science, Office of High Energy Physics, under Award Number DE-SC0011632
and by the the Walter Burke Institute for Theoretical Physics at Caltech.
HO is also supported in part by the Simons Investigator Award (MP-SIP-00005259) and by JSPS Grants-in-Aid for Scientific Research 23K03379. 
His work was performed in part at the Kavli Institute for the Physics and Mathematics of the Universe at the University of Tokyo, which is supported by the World Premier International Research Center Initiative, MEXT, Japan, at the Kavli Institute for Theoretical Physics (KITP) at the University of California, Santa Barbara, which is supported by grant NSF PHY-2309135, and at the Aspen Center for Physics, which is supported
by NSF grant PHY-1607611. 
YK is also supported by the INAMORI Frontier Program at Kyushu University and JSPS KAKENHI Grant Number 23K20046.
The authors thank Kyushu University Institute for Advanced Study and RIKEN Interdisciplinary Theoretical and Mathematical Sciences Program. Discussions during the “Kyushu IAS-iTHEMS workshop: Non-perturbative methods in QFT" were useful in completing this work.

\bibliographystyle{JHEP}
\bibliography{main.bib}

\end{document}